\documentstyle[emulateapj,psfig]{article}
\lefthead{Vignali et al.}
\righthead{High-Redshift Radio-Quiet Quasars}

\begin{document}

\slugcomment{Accepted for publication in 
{\em The Astrophysical Journal}}

\title{Probing the hard X--ray properties of high-redshift Radio-Quiet 
Quasars with ASCA}
\author{
C.~Vignali\altaffilmark{1} 
\authoremail{l\_vignali@kennet.bo.astro.it}, 
A.~Comastri\altaffilmark{2} 
\authoremail{comastri@astbo3.bo.astro.it}, 
M.~Cappi\altaffilmark{3,4} 
\authoremail{mcappi@tesre.bo.cnr.it}, 
G.G.C.~Palumbo\altaffilmark{1,3} 
\authoremail{ggcpalumbo@astbo3.bo.astro.it}, 
M.~Matsuoka\altaffilmark{4} 
\authoremail{matsuoka@postman.riken.go.jp}, 
and H.~Kubo\altaffilmark{4}
\authoremail{kubo@postman.riken.go.jp}
} 

\altaffiltext{1}{Dipartimento di Astronomia, Universit\`a di Bologna, 
via Zamboni 33, I-40126 Bologna, Italy}
\altaffiltext{2}{Osservatorio Astronomico di Bologna, via Zamboni 33, I-40126 
Bologna, Italy}
\altaffiltext{3}{Istituto per le Tecnologie e Studio delle Radiazioni Extraterrestri, 
ITeSRE/CNR, via Gobetti 101, I-40129 Bologna, Italy}
\altaffiltext{4}{The Institute of Physical and Chemical Research (RIKEN), 
2-1, Hirosawa, Wako, Saitama 351-01, Japan}
\authoremail{vignali@kennet.bo.astro.it}

\begin{abstract}
\noindent
This paper reports the X-ray spectral analysis of 5 high-redshift (z$\ge$2) 
radio-quiet quasars observed by ASCA. A simple power law continuum plus 
cold Galactic absorption model well fits all the spectra (tipically 
between $\sim$ 2--30 keV in the sources frame). 
Neither the X--ray spectral hardening, attributed to a reflection 
component and observed in Seyfert galaxies, nor the excess 
absorption previously detected in high-redshift radio-loud quasars, have been revealed. 
Only a marginal evidence of a neutral or mildly ionized FeK$\alpha$ line 
is found in one of the quasars. 
The average spectral slope in the observed 0.7--10 keV energy range, 
$<\Gamma>$ = 1.67$\pm{0.11}$ (dispersion $\sigma$ $\sim$ 0.07), 
appears to be flatter than that of low-z radio-quiet quasars 
($\Gamma$ $\simeq$ 1.9--2) and slightly steeper, 
but consistent with $\Gamma$ = 1.61$\pm{0.04}$ ($\sigma$ $\sim$ 0.10) 
of high-z radio-loud quasars. 
\end{abstract}

\section{Introduction}
\noindent
Quasars are well known to be powerful sources over almost the entire 
electromagnetic spectrum. In particular, spectra in the X--ray band play a remarkable 
role in providing vital clues to the understanding of their physical properties. 
Quasars, in fact, emit a noticeable fraction of their bolometric luminosity 
in this energy range and their X--ray emission appears to be closely associated with 
the activity induced by the advocated central compact object, most likely a massive black hole 
(see Mushotzky, Done \& Pounds 1993 for a review). X--ray spectra thus represent 
a powerful diagnostic probe of both the quasar environment and the radiation emission mechanism.\\
\noindent 
So far, X--ray observations have been mainly carried out for low-redshift quasars, both radio-quiet (RQQs) 
and radio-loud (RLQs). 
Spectral studies of low-z RQQs in the medium energy range 
have been performed both by EXOSAT in the 0.1--10 keV band (Comastri et al. 1992; Lawson et al. 1992) 
and Ginga in the 2--20 keV band (Williams et al. 1992). For RQQs 
the average 2--10 keV EXOSAT spectrum is similar to the Ginga one 
($<\Gamma>$ = 1.90$\pm{0.11}$ and $<\Gamma>$ = 2.03$\pm{0.16}$, respectively), with the RLQs showing a flatter 
X--ray spectral slope.  
It must be noted, however, that somewhat different spectral results have been obtained 
by Lawson and Turner (1997) through a re-analysis of 50 archival quasars observed by Ginga. 
They found $<\Gamma>_{\rm RQ}$ $\sim$ $<\Gamma>_{\rm RL}$ 
$\sim$ 1.7--1.8 for their full sample. 
This result may be partly explained if some RQQs of their sample have a 
more complex spectrum than a simple power law. 
For instance, the presence of a warm absorber and/or 
a reflection component could affect, if neglected (or if not properly modeled), 
the measure of the spectral slope.\\
\noindent 
In the softer energy range RQQs spectral slopes spread from $<\Gamma>$ $\simeq$ 2 in the 
0.2--3.5 keV Einstein IPC band (Wilkes \& Elvis 1987; Canizares \& White 1989) 
to $<\Gamma>$ $\sim$ 2.5 (and a wider dispersion of the indices) 
in the 0.1--2.4 keV ROSAT PSPC energy range (Brinkmann et al. 1992; 
Brunner et al. 1992; Laor et al. 1997). 
This spectral slope steepening towards softer energies has been interpreted as an 
additional soft component (Comastri et al. 1992; Mushotzky et al. 1993), similar to 
what has been found in Seyfert 1 galaxies (Turner \& Pounds 1989). \\
\noindent
Low-z RLQs are found to have sistematically flatter 
spectral indices than RQQs over the entire X--ray range, by $\Delta\Gamma$ $\sim$ 0.3 
in the 2--10 keV band (e.g. with EXOSAT, Lawson et al. 1992; with Ginga, Williams et al. 1992; with ASCA, Reeves et al. 1997) 
and $\Delta\Gamma$ $\sim$ 0.5 in the Einstein IPC energy range (Kriss \& Canizares 1985; Wilkes \& Elvis 1987). 
These spectral properties 
may suggest a different emission mechanism or an enhanced X--ray emission 
in RLQs perhaps due to beaming effects (Wilkes \& Elvis 1987).\\
\noindent 
The results obtained so far for low-z RQQs indicate that their 2--10 
keV power law spectral slopes are similar to those of low luminosity Seyfert 
1 galaxies ($\Gamma$ $\simeq$ 1.9-2.0, Nandra \& Pounds 1994; Nandra et al. 1997a). 
While the reflection ``hump", i.e. the typical sign of reprocessing gas, 
has not been clearly stated for RQQs, 
some evidence of ionized reprocessor (e.g. a ionized iron line) 
has been observed by ASCA in some RQQs, e.g. PG~1116+215 
(Nandra et al. 1996) and E~1821+643 (Kii et al. 1991; Yamashita et al. 1997), 
in the 10$^{45}$--10$^{46}$ erg s$^{-1}$ luminosity range. 
For instance, 
Nandra et al. (1997b) noticed that for high-luminosity quasars the iron line profile 
and intensity appear to be 
strongly luminosity-dependent. This result supports the inverse correlation between 
luminosity and Fe K emission line equivalent width previously 
reported by Iwasawa \& Taniguchi (1993).\\
\noindent 
On the other hand, the X--ray spectral properties of high-z RQQs 
are poorly known, in contrast with the more widely studied high-z RLQs 
(Elvis et al. 1994; Siebert et al. 1996; Cappi et al. 1997). 
This is due to the fact that in the X--ray band RLQs are about a factor 3 brighter 
than RQQs (Zamorani et al. 1981; Worrall et al. 1987). 
ROSAT observations indicate X--ray spectral slopes of high-z RQQs to be steeper than those of 
RLQs (Bechtold et al. 1994). Such X--ray spectra, however, are very poorly constrained. 
In a sample of 286 radio-quiet objects Fiore et al. (1998) 
found no clear indication of absorption. However most spectral properties 
have been derived from ``color" techniques and there are only 12 RQQs with z $>$ 2.2 in their sample.\\
\noindent 
Since RQQs represent the majority ($\sim$ 80--90\%, Osterbrock 1991) 
of the quasar population, a better knowledge of 
their X--ray spectral properties is of paramount importance in understanding 
their contribution to the X--ray background (XRB). ROSAT surveys (Hasinger et al. 1993) and 
theoretical models (Comastri et al. 1995) suggest the XRB may be accounted for by a 
large population of high-luminosity ($L_{\rm X}$ $\simeq$ 10$^{44}$ erg s$^{-1}$), 
high-redshift absorbed ($N_{\rm H}$ $\simeq$ 10$^{22}$--10$^{23}$ cm$^{-2}$) 
RQQs, with spectral slopes similar to the low-z population ones. \\
\noindent 
Given the data and hypothesis listed above, an ASCA program has been started 
in order to study spectral and evolutionary 
properties of a sample of high-redshift RQQs. 
The aim of this program was to obtain 
{\bf a)} reliable measurements of RQQs X--ray spectra from which 
to deduce X--ray slopes and absorption at high-redshift, 
{\bf b)} compare these data with the available low-z RQQs and high-z RLQs data 
and {\bf c)} search for complex structures, i.e. 
K$\alpha$ emission line and reflection component. 
$H_{0}$ = 50 km s$^{-1}$ Mpc$^{-1}$ and $q_{0}$ = 0.5 are assumed throughout. 

\section{Observations and Data Analysis}

\subsection{The Sample}
\noindent
In this paper, results obtained from ASCA observations of a sample of 8 high-redshift ($z\geq1.9$) quasars 
are presented (see Table~1). 
The sample was obtained cross-correlating 
the V\'eron-V\'eron quasars catalogue (V\'eron-Cetty and V\'eron 1996) 
with the ROSAT All-Sky Survey catalogue of X-ray sources (Voges et al. 1996). 
All the selected quasars have a ROSAT count rate $\gtrsim$ 0.04 counts s$^{-1}$. 
Even if not complete and probably biased toward less absorbed sources, 
this sample is adequate for this study as the main scope is to obtain, for the first time, 
a reliable measurement of the X--ray spectral properties 
of a sample of high-z RQQs choosen amongst the brightest and highest redshift 
sources. 
5 clear detections and 3 upper limits have been obtained so far 
with the ASCA X--ray telescope and 5 more objects are approved for observation in the ongoing AO6. 
For three objects also ROSAT PSPC archival data have been analyzed. 

\placetable{tab1}

\subsection{ASCA Data Reduction}
\noindent
The high-z RQQs were observed by the ASCA satellite (Tanaka, Inoue \& Holt 1994) 
during the AO4 and AO5 phases, according to the observation log shown in Table~2. 
The focal plane instruments consist of two solid-state imaging spectrometers (SIS, 
Gendreau 1995) and two gas scintillation imaging spectrometers 
(GIS, Makishima et al. 1996), which provide a good 
spectral resolution (nominal FWHM $\sim$ 2\% and $\sim$ 8\% at 5.9 keV, 
respectively, when the satellite was launched) 
and broad band ($\sim$ 0.5--10 keV) capabilities. 
The observations were performed in FAINT mode and then corrected 
for dark frame error and echo uncertainties (Otani \& Dotani 1994). 
The data were screened with the version 1.3 of the 
{\sc XSELECT} package, in order to include only data collected when: 
the spacecraft was outside the South Atlantic Anomaly; the radiation belt monitor 
was less than 100 counts s$^{-1}$; the magnetic cut-off rigidity was greater than 
6 GeV c$^{-1}$ for SIS and 4 GeV c$^{-1}$ for GIS; the elevation angle above the Earth limb 
greater than 5$^{\circ}$ and the angle above the Sun-illuminated Earth limb 
greater than 15$^{\circ}$. ``Hot" and flickering pixels were removed from the SIS 
using the standard algorithm. 
SIS grades 0, 2, 3 and 4 were considered in the data reduction. 
For the GIS data the recently available method of rejecting ``hard particle flares" 
using the so-called HO2 count rate, was employed, as well as the standard ``rise-time" criteria. 
Source counts were extracted from circles of about 
6$\arcmin$ radius for GIS and 3$\arcmin$ for SIS centered on the source, and background spectra 
were extracted from source-free regions from the same CCD chip for SIS and for the same 
FOV for GIS. 
Relevant data are reported in Table~2. 
For the SIS datasets, appropriate detector redistribution matrices were generated using the ftool 
sisrmg (v1.1); for the GIS data the 1995 response matrices gis2v4\_0.rmf and gis3v4\_0.rmf were used. 
Ancillary response files were created for all detectors with the ftool ascaarf (v2.72). 
Spectral analysis has been carried out with version 10.0 of the {\sc XSPEC} program 
(Arnaud 1996). 

\placetable{tab2} 

\subsection{ASCA Spectral Analysis and Main Results}
\noindent
GIS and SIS spectra were binned with more than 20 counts per bin in order to apply 
$\chi^2$ statistics. 
The statistical quality of the GIS data for 1101$-$264 does not allow to 
extract a useful spectrum for this object. For all the other quasars a good agreement, 
within the errors, has been found between SIS and GIS data. The spectral analysis 
has been carried out by simultaneously fitting SIS and GIS spectra, allowing the 
relative normalizations free to vary in order to account for 
residual discrepancies in the absolute flux calibration. 
Throughout the paper, errors are given at 90\% 
confidence level for one interesting parameter ($\Delta\chi^2$ = 2.71, Avni 1976). 
X--ray spectral fits were performed using a single absorbed power law model 
plus photoelectric absorption from cold material assuming cosmic abundances 
(Anders \& Grevesse 1989) and cross sections derived from Balucinska-Church \& McCammon 
(1992). At first the column density $N_{\rm H}$ was fixed to the Galactic value (Dickey \& Lockman 1990), 
then it was left free to vary. 
The resulting best-fit parameters are showed in Table~3. 

\placetable{tab3}

\noindent 
Since the sources are rather faint, the effect of uncertainties of the background 
subtraction on the obtained parameters has been checked by varying its normalization 
by a $+$10\% and $-$10\%. 
The results of this test give confidence that the spectral properties weakly depend 
on the assumed background normalizations. \\
\noindent 
The 0.7--10 keV (observer frame) average slope is 
$<\Gamma>$ = 1.67$\pm{0.11}$ (dispersion $\sigma$ $\sim$ 0.07) when 
the absorption is fixed to the Galactic value and $<\Gamma>$ = 1.74$\pm{0.17}$ 
($\sigma$ $\sim$ 0.18) when $N_{\rm H}$ is left free to vary. 
The best-fit spectra and the two-dimensional $\chi^2$ contour plots in the $N_{\rm H}$-$\Gamma$ space parameters 
are shown in Fig.~1. 
In all cases the Galactic $N_{\rm H}$ is consistent with the data, 
even if values greater than the Galactic one cannot be completely ruled out. 
It must be also stressed that the SIS detectors suffer from a calibration problem at 
low energy, which introduces a systematic error in the determination of $N_{\rm H}$ 
of the order of 2-3 $\times$ 10$^{20}$ cm$^{-2}$ (Cappi et al. 1998). 
In order to minimize the effects of this calibration problem on the spectral analysis 
we have not considered SIS data for energies $<$ 0.7 keV.

\placefigure{fig1}
\notetoeditor{Fig.~1 consists of 10 postscript files, divided into 2 columns}

\noindent
Spectral fittings with two separate absorbers, one at z = 0 fixed at the 
Galactic value and one at the quasar redshift (with $N_{\rm H}$ free to vary) 
were then repeated, but no improvement was found. 
The upper limits on $N_{{\rm H}_{\rm int}}$ 
are reported in Table~4. 
We have also searched for the presence of an iron emission line. There is 
only a very marginal ($\sim$ 95\% according to the F--test) detection of a 
FeK$\alpha$ line in the quasar 1101$-$264, characterized by a ``nominal" 
high equivalent width ($\sim$ 700 eV in the source rest-frame). 
The upper limits on the equivalent widths of the neutral and H-like lines are reported 
in Table~4. 
It should be noted that the obtained upper limits on the line equivalent 
widths may be slightly over-estimated when the line 
is weak or totally absent (Yaqoob 1998). 

\placetable{tab4}

\noindent
Given that all the objects are clustered around z $\simeq$ 2, have similar power law shapes and 
do not show evidence of strong intrinsic absorption, a co-added SIS 
spectrum has been computed rescaling 
all the objects at z = 2, in order to search for more complex features, expecially the high-energy 
``hump" and the iron emission line. The best-fit model has been achieved with a power law 
($\Gamma$ = 1.70$^{+0.04}_{-0.06}$) plus absorption consistent with the average Galactic value 
(Fig.~2). 

\placefigure{fig2}

\noindent
We have also simultaneously fitted SIS and GIS spectra of all the objects with the same model, 
allowing the relative normalizations free to vary, while the individual redshifts 
have been fixed for each object. 
The best-fit SIS+GIS slope, $\Gamma$ = 1.65$\pm{0.04}$, agrees very well with that derived 
from the co-added SIS spectrum.\\ 
The data have also been fitted with thermal models (e.g. Bremsstrahlung and Raymond-Smith), which 
sistematically yielded very high temperatures (from $\sim$ 10 to $\sim$ 40 keV 
in the sources frame), thus difficult to be physically interpreted. 
Given their significantly higher $\chi^{2}$ compared to a power law fit ($\Delta\chi^{2}$ $\sim$ 3--20), 
these models can be generally ruled out even when the abundances are left free to vary. 
Two-components models are not required. \\
\noindent 
Since the observed energy range of 0.7--10 keV corresponds 
to a range of $\sim$ 2--30 keV in the sources frame, it is well adequate to search for 
the typical hardening of the X--ray spectral slope due to the 
reflection component, as observed in Seyfert galaxies (Nandra \& Pounds 1994). 
This search has been done by performing 
separate spectral fits in the 0.7--3 keV and 3--10 keV observed energy ranges for all objects. 
In no case the resulting high-energy slope was significantly flatter than the low-energy one, 
although the upper limits on the possible flattening are not particularly strong ($\Delta\Gamma$ $<$ 0.2). 
In order to improve on this result, we have then applied a reflection component 
to both the co-added SIS spectrum and to the SIS+GIS one: 
for the ``average" reflection parameter R (defined 
as the ratio of the reflected to the direct component) values of 1.25$^{+1.53}_{-0.89}$ and 
0.76$^{+0.85}_{-0.58}$ have been found, respectively, with no significant improvement in the fit.
As a last test, the power law slope was fixed to the typical Seyfert 1 
photon index, $\Gamma$ $\sim$ 1.9 (Nandra \& Pounds 1994). The resulting best-fit 
reflection parameter is R = 2.30$^{+0.49}_{-0.46}$ for the 
SIS co-added spectrum and 
R = 2.49$^{+0.40}_{-0.37}$ for the SIS+GIS spectrum, but the $\chi^{2}$ value is significantly 
higher ($\Delta\chi^{2}$ $\simeq$ 8) than that obtained without any reflection 
component. 
The lack of significant reflection is also corroborated by the iron line upper limits, i.e. 
the value for R obtained with a power law slope fixed at $\Gamma$ = 1.9, as seen in Seyfert 1 galaxies, 
is high if compared to the tight upper limit obtained for the equivalent width 
of the iron emission line, which is of the order of 100 eV (source frame). 

\section{ROSAT PSPC Results}
\noindent
The available archival ROSAT PSPC pointed observations for three quasars have been analyzed 
and the results are reported in Table~5. 
A factor $\sim$ 2 flux variability in a time interval of about 
30 months (observer frame) has been revealed for 1101$-$264 between ROSAT and ASCA observations, 
but no spectral variability has been detected. 
The PSPC spectra are well described by a power law model without excess absorption, thus confirming 
ASCA results. For 1101$-$264 the ROSAT and ASCA slopes are consistent with each other, while for the 
other two objects the ROSAT slope is significantly steeper (Fig.~3). 
This result may possibly be ascribed in part to the large off-axis angle at which 
0040$+$0034 has been observed or possibly to a somewhat softer component 
which went unrevealed by ASCA (e.g. a thermal component with kT $\sim$ 130--350 eV rest-frame) or 
even 
to remaining cross-calibration uncertainties. 

\placetable{tab5}

\placefigure{fig3}

\section{Discussion}
\noindent
Before drawing firm conclusions from the following arguments one should bear in mind 
that most of the results discussed from both the present work and published 
papers are based on small numbers of objects, poor statistics and no sample considered 
is anywhere near a statistically complete sample. Having said this, it is nevertheless 
reasonable to attempt a physical interpretation of the present results.\\ 
\noindent 
From the present ASCA observations of high-redshift RQQs 
there is no evidence of any intrinsic absorption by neutral material. 
It must be noted, however, that our sample suffers from its selection criteria, 
which may produce a bias towards mildly or not- absorbed sources (see $\oint$ 2.1). 
If confirmed by future observations, this result would support 
a scenario in which RLQs and RQQs evolve differently. 
In fact, Elvis et al. (1994) have shown that low-energy cutoffs due to X--ray 
absorption are common in ROSAT spectra of high-z RLQs, with typical column densities 
of the order of 10$^{21}$--10$^{22}$ cm$^{-2}$, depending on the location of 
the absorber. A more complete analysis has been carried out by Cappi et al. (1997) 
on ASCA and ROSAT data. 
Their results favour, at least in 2 of the 6 absorbed objects (out of 9), 
an intrinsic origin for the absorption and a strong clustering 
of the continuum spectral slopes in the $\sim$ 1--40 keV (quasar-frame) range around 
$\Gamma$ $\sim$ 1.5--1.6 with a small dispersion. 
A recent study has been carried out by Fiore et al. (1998) on 
a much larger sample of about 500 quasars 
(0.1 $<$ z $<$ 4.1) found in the WGA (White, Giommi \& Angelini 1994) 
and ROSATSRC catalogues of pointed ROSAT PSPC 
observations (Voges et al. 1994). The results confirm 
that low-energy absorption is more common (and likely exclusive) in RLQs 
rather than in RQQs (see also Laor et al. 1997; Yuan et al. 1998). 
This rules out intervening systems as the principle cause of the excess absorption, 
since this would affect both the RLQs and the RQQs X--ray spectra, which is not the case. 
Moreover, recent ASCA results (Reeves et al. 1997) indicate that 
the column density itself is 
strongly correlated with redshift (i.e. more distant radio-loud objects suffer 
from greater absorption). These evidences imply an evolution with cosmic time of the absorbers 
in RLQs (Elvis 1996).\\
%
%
\noindent 
The most important result from the present analysis is that all quasars have 
similar slopes, with an average value of $<\Gamma>$ = 1.67$\pm{0.11}$ 
and a small dispersion ($\sigma$ $\sim$ 0.07). 
The source frame energy band of the present sample 
($\sim$ 2 -- 33 keV) compares well with Lawson and Turner's ($\sim$ 2 -- 35 keV), and is slightly 
harder than Reeves et al.'s (1997) sample ($\sim$ 0.5 -- 23 keV). 
The spectral results derived from the present sample seem to 
indicate flatter X--ray spectra for high-z RQQs 
than for low-z objects (Reeves et al. 1997; Lawson \& Turner 1997), as 
either we are sampling the RQQs lower tail of the photon indices distribution (see Fig.~4) 
or the primary emission mechanism is really different at high redshift, thus suggesting 
an evolutionary scenario for the RQQs population. 

\placefigure{fig4}

\noindent
Although no significant correlation for the quasars parameters ($\Gamma$ - $L_{\rm X}$ - z) can be derived 
because of poor statistics and the small number of objects of the present sample, 
it is however remarkable that the present analysis extends considerably the 
redshift range of radio-quiet quasars with respect to the past (Fig.~5). 

\placefigure{fig5}

\noindent
The average slope of the present sample, $<\Gamma>$ = 1.67, is steeper than 
that of the XRB, $\Gamma$ $\simeq$ 1.4 (Marshall et al. 1980; Gendreau et al. 1995). 
The obtained steep slope, together with the absence of excess absorption, suggests 
that these quasars do not contribute significantly to the XRB. 
Larger samples, preferably of hard X--ray selected 
objects, are needed, however, to test theoretical model such as in Comastri et al. (1995). \\
\noindent 
ASCA data have not revealed the presence of a reflection component, due to 
neutral or ionized matter surrounding the source. 
The general lack of reflection features could in principle be related 
to the quasars accretion rate: if they are accreting close to their 
Eddington limit, this can lead to an optically thin disc and no reflection would be seen. 
Alternatively, the iron in the disc may be completely ionized (\.{Z}ycki \& Czerny 1994) 
and the lighter elements fully stripped of electrons, which would smooth out any reflection feature. 
In fact, as suggested by Nandra et al. (1998b) for high-luminosity quasars, 
the Compton reflection could be seen without suffering strong absorption in the disc. 
The main consequence consists of a less evident 
contrast between the reflection component and the underlying continuum, thus the former 
should appear as a weak reflection ``hump" in X--ray spectra. 
An alternative possibility is that the lack of a clear reflection component and iron 
line may also indicate either 
different environmental or geometrical properties for RQQs with respect to Seyfert galaxies. 
Observations of 1101$-$264 with AXAF and XMM would clearly help to shed light on the 
reality of the iron line in this source. 
A more detailed analysis will be possible 
with the larger RQQs sample soon available.

\section{Conclusions}
\noindent
From the X--ray spectral analysis of 5 high-redshift RQQs observed with the ASCA satellite the 
following main results have been derived:\\
$\bullet$ the high-luminosity ($L_{\rm X}$ $\approx$ 10$^{46}$ erg s$^{-1}$) 
objects of the present sample do not show imprints of 
cold matter either in transmission (absorption) or in reflection 
(Fe K$\alpha$ emission line and Compton ``hump");\\
$\bullet$ the different spectral slopes between high-z RQQs and low-z RQQs may suggest 
a different emission mechanism, but a much more improved statistics is 
required before drawing firm conclusions;\\
$\bullet$ the average X--ray spectrum of the present high-z RQQs sample ($<\Gamma>$ = 1.67$\pm{0.11}$ 
in the $\sim$ 2--30 keV source frame energy range) is consistent with the high-z RLQs one;\\
$\bullet$ it is possible that RQQs and RLQs follow a different evolution, i.e. 
they have different environments at early epochs. 

\begin{acknowledgements}
\noindent 
We thank the ASCA team who operate the satellite and 
maintain the software and database. 
This work has made use of the NASA/IPAC Extragalactic Database (NED) 
which is operated by the Jet Propulsion Laboratory, Caltech, under contract 
with the National Aereonautics and Space Administration, of data obtained 
through the High Energy Astrophysics Science Archive Research (HEASARC) 
Center Online Service, provided 
by the Goddard Space Flight Center, and of the Simbad database, operated at CDS, Strasbourg, 
France.
Financial support from Italian Space Agency under the contract ARS--96--70 
and MURST is acknowledged by CV, AC, MC and GGCP. 
The authors would also like to thank G. Zamorani for scientific advices and 
a careful reading of the manuscript and the anonymous referee for his/her comments.
\end{acknowledgements}

\newpage

\figcaption[]
{Best-fit spectra and contour plots of the ASCA data, fitted with a single power law model. The contours represent 
the 68, 90 and 99\% confidence contours for the parameters $N_{\rm H}$ -- $\Gamma$ 
for SIS (solid line) 
and GIS (dashed line). The left margin of the figure represents the Galactic absorption (from Dickey \& 
Lockman 1990).
\label{fig1}}

\figcaption[]
{Plot of the co-added spectrum obtained re-scaling the energy scale of each quasar to z = 2 (left panel). 
68, 90 and 99\% confidence contours in the $N_{\rm H}$ -- $\Gamma$ plane obtained from the same spectrum 
(right panel). 
\label{fig2}}

\figcaption[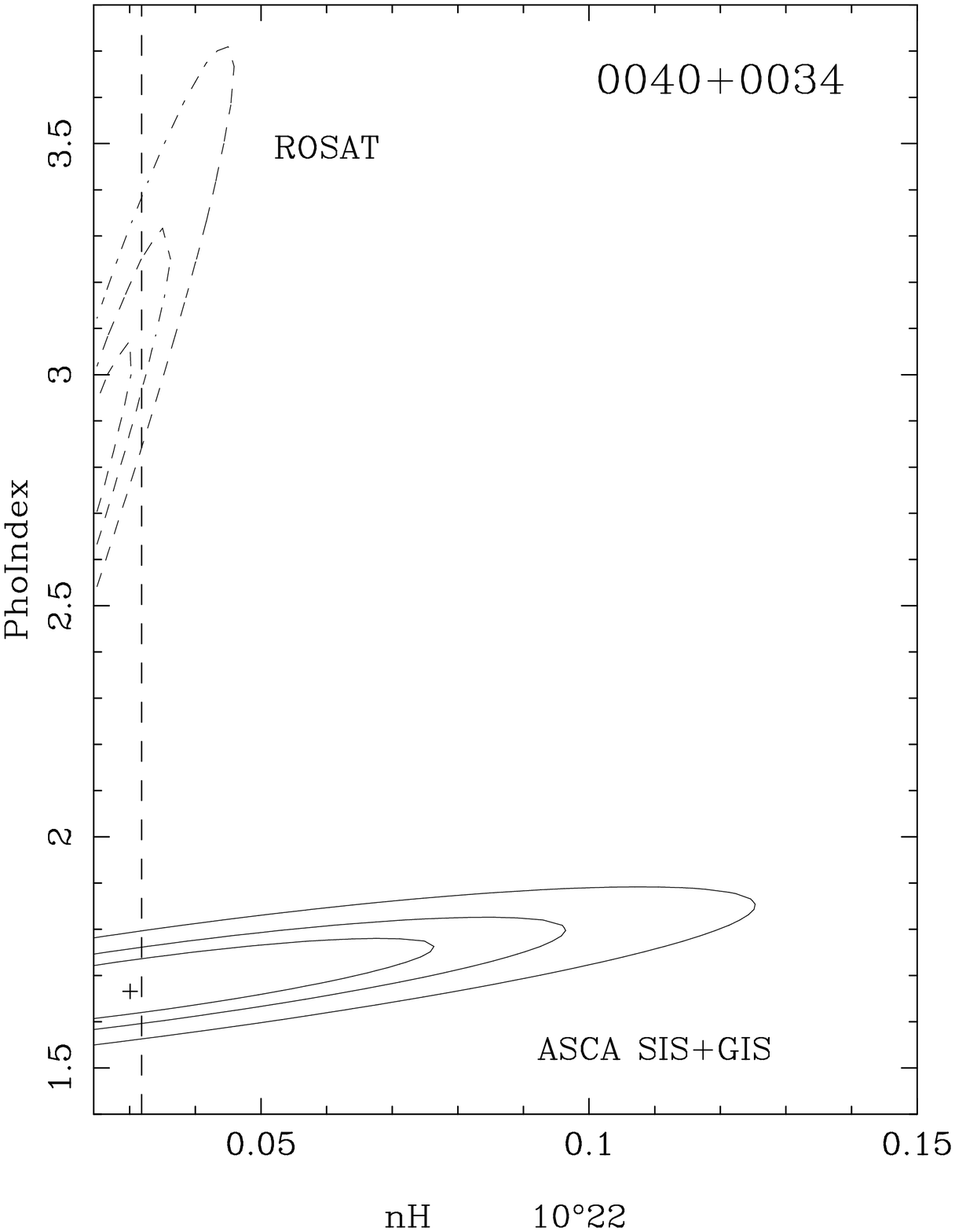,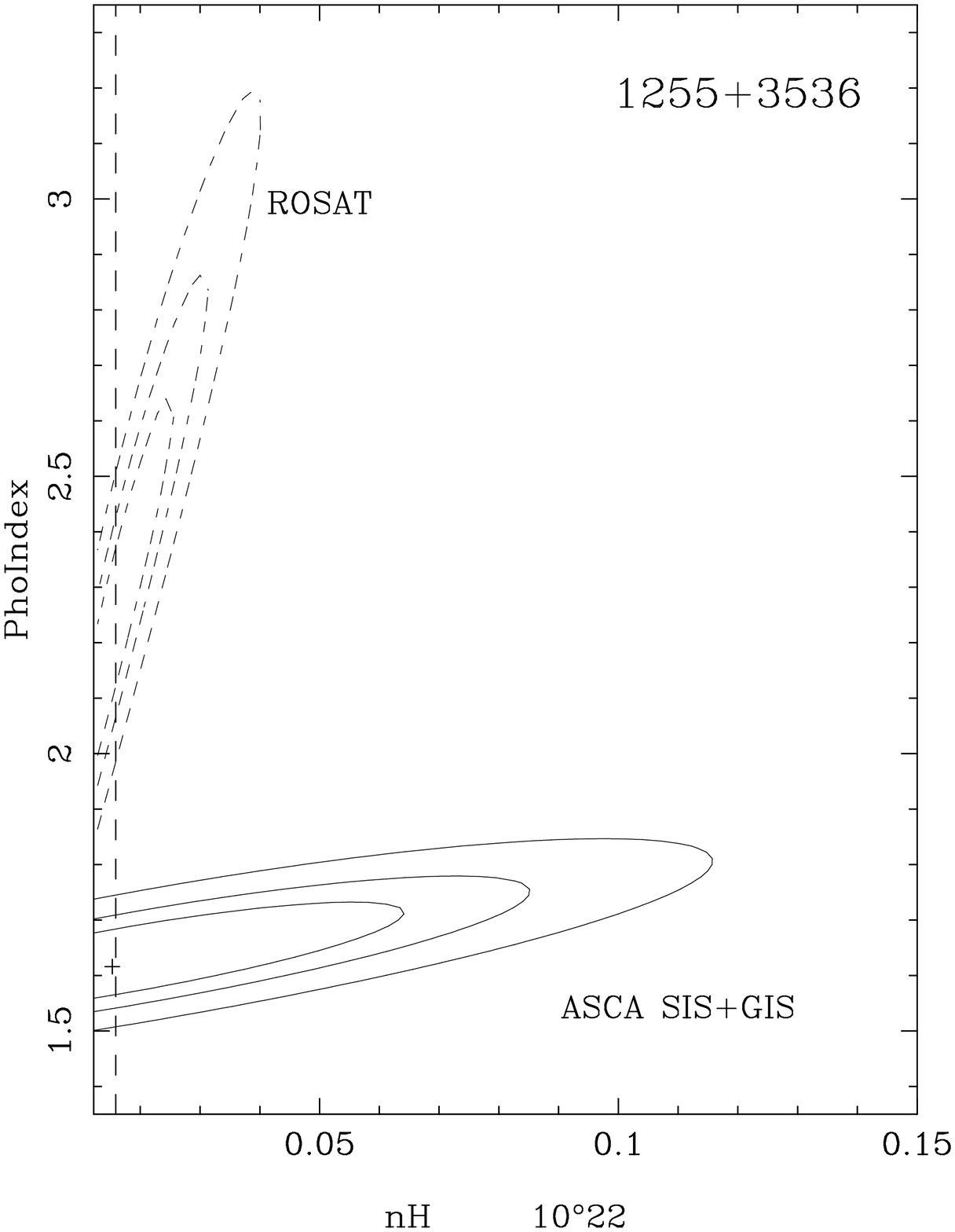]
{Comparison of ASCA and ROSAT 68, 90 and 99\% $\Gamma$ -- $N_{\rm H}$ 
confidence contours for 0040$+$0034 and 1255$+$3536. The dashed line represents the 
error on the Galactic absorption.
\label{fig3}}

\figcaption[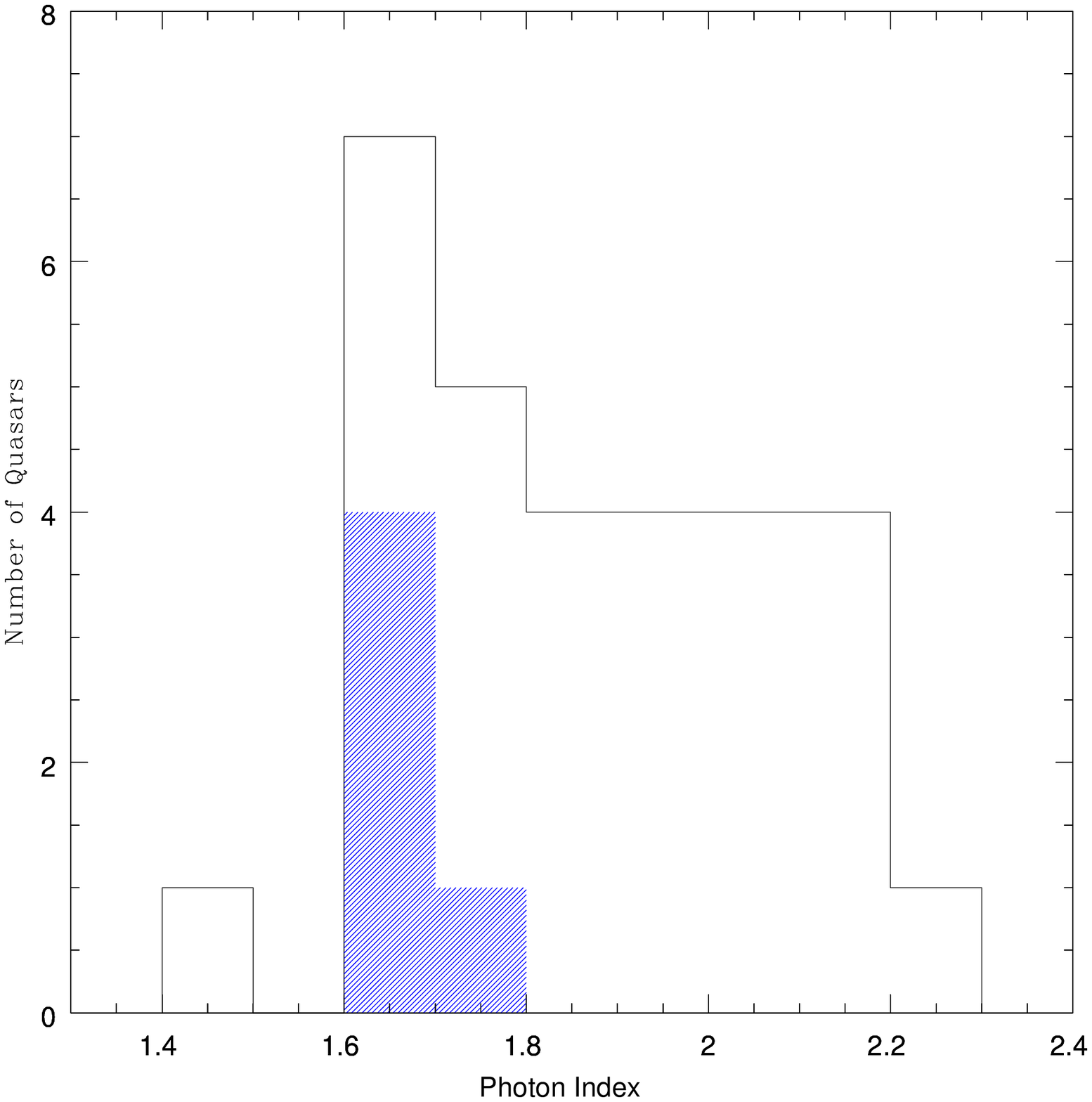]
{A histogram distribution of RQQs 
X--ray photon indices for the present sample (shaded area) compared to 
Lawson \& Turner 1997 Ginga sample, Reeves et al. 1997 ASCA sample and EXOSAT 
one (Comastri et al. 1992, Lawson et al. 1992). 
When the same object is present both in Ginga and ASCA samples, the best-fit (lower $\chi^{2}$) results 
have been choosen. 
\label{fig4}}

\figcaption[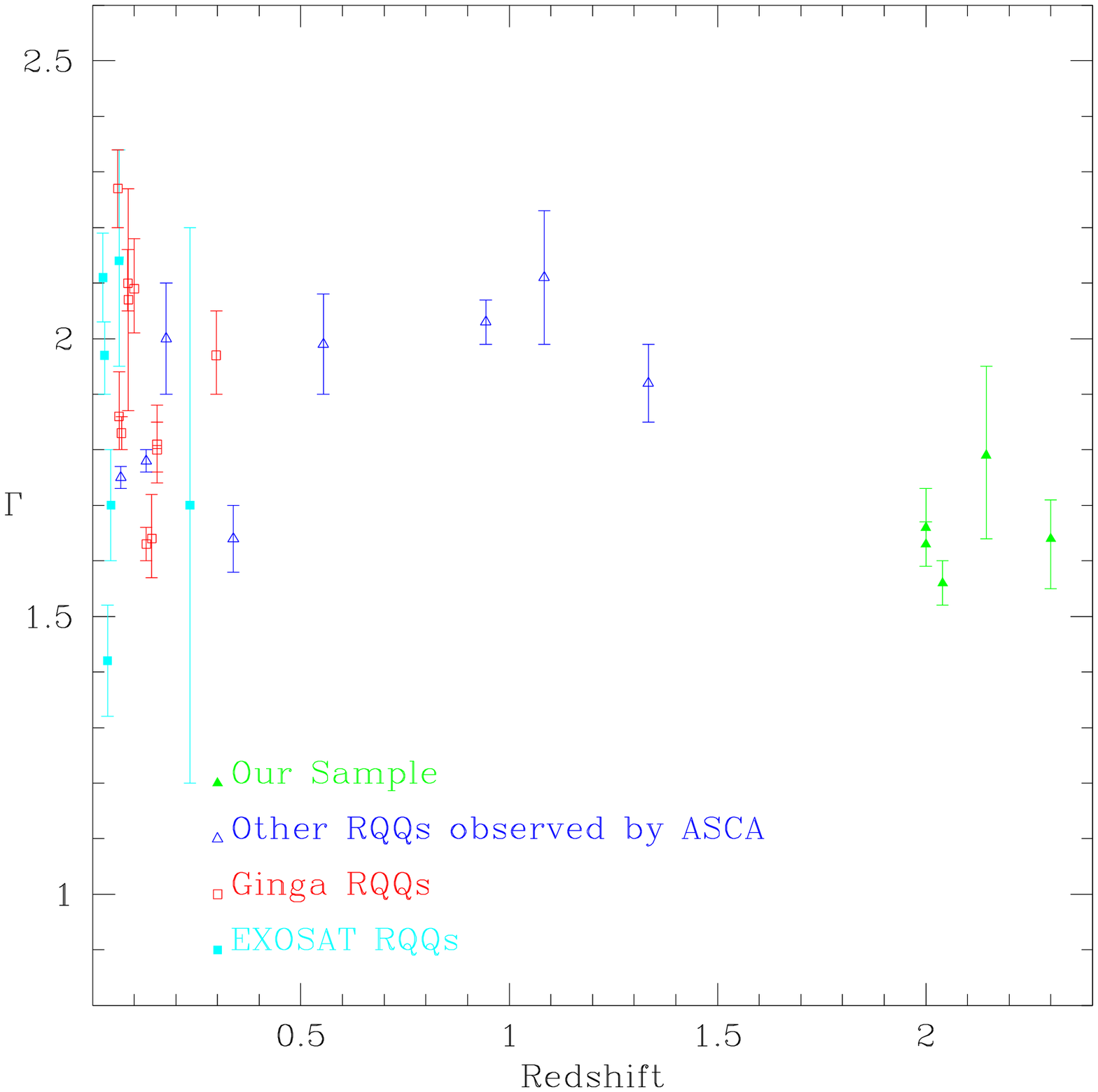]
{Plot of redshift versus photon index for the present sample, Reeves et al. 1997 ASCA RQQs subsample, 
for the Ginga one by Lawson \& Turner 1997 and for EXOSAT objects (Lawson et al. 1992, Comastri et al. 1992). 
When the same object is present both in Ginga and ASCA samples, the best-fit (lower $\chi^{2}$) results 
have been choosen. 
\label{fig5}}

\newpage
\begin{deluxetable}{ccccccc}
\small
\tablecaption{The Radio-Quiet Quasar Sample. \label{tab1}}
\tablewidth{0pt}
\tablehead
{
\colhead{Object} & \colhead{R.A.}   & \colhead{Decl.}   & \colhead{z} &
\colhead{$N_{{\rm H}_{\rm gal}}$\tablenotemark{a}}  & \colhead{$m_{\rm V}$} & \colhead{$R_{\rm L}$\tablenotemark{b}} \\
\colhead{} & \colhead{(J2000)} & \colhead{(J2000)} & \colhead{} & 
\colhead{} & \colhead{} & \colhead{} \nl
}
\startdata
0040$+$0034 (UM~269) & 00$^{\rm h}$43$^{\rm m}$19.7$^{\rm s}$ & 
$+$00$\arcdeg$51$\arcmin$16$\arcsec$&2.00 & 2.45 & 18.0 & $<$ 0.67 \nl
0300$-$4342 (C~25.36) & 03 01 55.7 & $-$43 30 40 & 2.30 & 1.83 & 
19.2 & $<$ 1.15 \nl
1101$-$264 (PG~1101$-$264) & 11 03 25.2 & $-$26 45 15.2 & 2.15 & 
5.68 & 16.0 & $<$ -0.22 \nl
1255$+$3536 (WEE~83) & 12 58 09 & $+$35 19 48 & 2.04 & 
1.22 & 20.6 & $<$ 1.71 \nl
1352$-$2242 (CTS~327) & 13 55 43.4 & $-$22 57 07 & 2.00 & 
5.88 & 18.2 & $<$ 0.72 \nl
1559$+$089 & 16 02 22.5 & $+$08 45 36 & 2.27 & 3.84 & 16.7 & $<$ 1.00 \nl
1725$+$503 & 17 26 57.5 & $+$50 15 48 & 2.10 & 2.58 & 20.4 & $<$ 1.63 \nl
1726$+$504 & 17 27 38.9 & $+$50 26 06 & 1.90 & 2.58 & 19.9 & $<$ 1.43 \nl
\enddata
\tablenotetext{a}{In units of 10$^{20}$ cm$^{-2}$, Dickey \& Lockman 1990}
\tablenotetext{b}{Radio loudness, defined as $R_{\rm L}$ = Log (f$_{\rm 5 GHz}$/f$_{\rm V}$)}
\end{deluxetable}

\newpage
\begin{deluxetable}{cccccccc}
\small
\tablecaption{ASCA Observation Log. \label{tab2}}
\tablewidth{0pt}
\tablehead
{
\colhead{} & \colhead{} & \multicolumn{2}{c}{Exposure\tablenotemark{a}} &
\colhead{} & \multicolumn{2}{c}{Count Rate\tablenotemark{a}} \\
\colhead{} & \colhead{} & \multicolumn{2}{c}{(ks)} &
\colhead{} & \multicolumn{2}{c}{($\times$ 10$^{-2}$ c s$^{-1}$)} \\
\cline{3-4} \cline{6-7} \\
\colhead{Object} & \colhead{Date} & \colhead{GIS} & \colhead{SIS} & 
\colhead{} & \colhead{GIS} & \colhead{SIS} 
}
\startdata
0040$+$0034 & 1997 Jul 13 & 36.0 & 29.9 & & 3.25 & 3.82 \nl
0300$-$4342 & 1997 Jan 23 & 37.5 & 37.8 & & 1.17 & 1.31 \nl
1101$-$264  & 1996 Jun 16 & 19.7 & 17.4 & & 0.56 & 0.98 \nl
1255$+$3536 & 1997 Jun 23 & 36.7 & 36.5 & & 2.00 & 3.44 \nl
1352$-$2242 & 1997 Feb 02 & 31.6 & 31.1 & & 1.50 & 1.68 \nl
1559$+$089  & 1996 Aug 23 & 18.7 & 17.8 & & 0.17 & 0.22 \nl
1725$+$503  & 1995 Mar 04 & 39.0 & 41.2 & & 0.33 & 0.45 \nl
1726$+$504  & 1995 Mar 04 & 39.0 & 41.2 & & 0.27 & 0.36 \nl
\enddata
\tablecomments{All the observations were carried out in 1-CCD mode, except 
for 1725$+$503 and 1726$+$504 (2-CCD mode)}
\tablenotetext{a}{The reported values for the GIS and SIS are averaged over 
the detectors (GIS2 with GIS3 and SIS0 with SIS1)}
\end{deluxetable}

\newpage

\begin{deluxetable}{ccccccc}
\scriptsize
\tablecaption{SIS + GIS X--ray Spectral Fits.
\label{tab3}}
\tablewidth{0pt}
\tablehead
{
\colhead{Object} & \colhead{$N_{\rm H}$} & \colhead{$\Gamma$}
& Refl. & \colhead{$\chi^{2}$/dof} & \colhead{$F_{2-10 keV}$\tablenotemark{a}} &
\colhead{$L_{2-10 keV}$\tablenotemark{b}} \\
\colhead{} & \colhead{(10$^{20}$ cm$^{-2}$)} & \colhead{} &
\colhead{} & \colhead{} & \colhead{} & \colhead{} \nl
}
\startdata
0040$+$0034 & $\equiv$$N_{{\rm H}_{\rm gal}}$ & 1.66$^{+0.07}_{-0.06}$ 
& & 240/237 & 15 & 2.7 \nl
 & $<$ 8.05 & 1.67$^{+0.12}_{-0.07}$ & & 239/236 & & \nl
 & $\equiv$$N_{{\rm H}_{\rm gal}}$ & 1.71$^{+0.18}_{-0.14}$ & $<$ 2.34 & 239/236 & & \nl
 & $\equiv$$N_{{\rm H}_{\rm gal}}$ & 1.90 fr. & 2.34$^{+0.63}_{-0.58}$ & 241/237 & & \nl
\tableline
0300$-$4342 & $\equiv$$N_{{\rm H}_{\rm gal}}$ & 1.64$^{+0.12}_{-0.15}$ & & 168/148 & 5.0 & 1.2\nl
 & $<$ 5.66 & 1.65$^{+0.12}_{-0.16}$ & & 168/147 & & \nl
 & $\equiv$$N_{{\rm H}_{\rm gal}}$ & 1.96$^{+0.19}_{-0.33}$ & 3.62$^{+6.38}_{-2.90}$ & 
164/147 & & \nl
 & $\equiv$$N_{{\rm H}_{\rm gal}}$ & 1.90 fr. & 3.04$^{+1.49}_{-1.27}$ & 164/148 & & \nl
\tableline
1101$-$264 & $\equiv$$N_{{\rm H}_{\rm gal}}$ & 1.79$^{+0.26}_{-0.25}$ & & 29.3/23 
& 2.9 & 0.7 \nl
& $<$ 45.5 & 2.06$^{+0.73}_{-0.45}$ & & 27.9/22 & & \nl
& $\equiv$$N_{{\rm H}_{\rm gal}}$ & 1.70$^{+1.69}_{-0.23}$ & unc. & 29.2/22 & & \nl
& $\equiv$$N_{{\rm H}_{\rm gal}}$ & 1.90 fr. & $<$ 3.61 & 29.6/23 & & \nl
\tableline
1255$+$3536 & $\equiv$$N_{{\rm H}_{\rm gal}}$ & 1.62$^{+0.06}_{-0.07}$ & & 187/198 & 11 & 2.1\nl
 & $<$ 6.83 & 1.62$^{+0.12}_{-0.07}$ & & 187/197 & & \nl
 & $\equiv$$N_{{\rm H}_{\rm gal}}$ & 1.56$^{+0.16}_{-0.08}$ & $<$ 1.07 & 187/197 & & \nl
 & $\equiv$$N_{{\rm H}_{\rm gal}}$ & 1.90 fr. & 2.65$^{+0.68}_{-0.63}$ & 196/198 & & \nl
\tableline
1352$-$2242 & $\equiv$$N_{{\rm H}_{\rm gal}}$ & 1.66$\pm{0.12}$ & & 145/147 & 6.7 & 1.1 \nl
 & $<$ 11.2 & 1.67$\pm{0.12}$ & & 145/146 & & \nl
 & $\equiv$$N_{{\rm H}_{\rm gal}}$ & 1.84$^{+0.80}_{-0.28}$ & $<$ 23 & 144/146 & & \nl
 & $\equiv$$N_{{\rm H}_{\rm gal}}$ & 1.90 fr. & 2.43$^{+1.18}_{-1.02}$ & 144/146 & & \nl
\tableline
\tableline
1559$+$089 & $\equiv$$N_{{\rm H}_{\rm gal}}$ & 1.8 & & & $<$ 0.6 & $<$ 0.16 \nl
1725$+$503 & $\equiv$$N_{{\rm H}_{\rm gal}}$ & 1.8 & & & $<$ 1.2 & $<$ 0.28 \nl
1726$+$504 & $\equiv$$N_{{\rm H}_{\rm gal}}$ & 1.8 & & & $<$ 1 & $<$ 0.19 \nl
\enddata
\tablenotetext{a}{Absorbed flux in the 2-10 (observer frame) energy range in units of 
10$^{-13}$ erg s$^{-1}$ cm$^{-2}$} 
\tablenotetext{b}{Intrinsic 2-10 keV (quasar frame) luminosity in units of 10$^{46}$ 
erg s$^{-1}$}
\tablecomments{Intervals are at 90\% confidence for one interesting parameter ($\Delta\chi^{2}$ = 2.71). 
For the last three quasars of the list the data have been obtained as 3 $\sigma$ upper limits and assuming 
Galactic absorption and $\Gamma$ = 1.8} 
\end{deluxetable}

\newpage
\begin{deluxetable}{cccc}
\small
\tablecaption{Upper limits on the intrinsic column density and iron line measurements
\label{tab4}}
\tablewidth{0pt}
\tablehead
{
\colhead{Object} & \colhead{$N_{{\rm H}_{\rm int}}$} & 
\colhead{EW$_{6.4 keV}$\tablenotemark{a}} & \colhead{EW$_{6.97 keV}$\tablenotemark{b}} \\
\colhead{} & \colhead{(10$^{21}$ cm$^{-2}$)} & \colhead{} & \colhead{} \nl
}
\startdata
0040$+$0034 & $<$ 8.75 & $<$ 90 & $<$ 96 \nl 
0300$-$4342 & $<$ 6.66 & $<$ 87 & $<$ 163 \nl
1101$-$264 & $<$ 49.6 & 692$^{+563}_{-562}$ & $<$ 786 \nl
1255$+$3536 & $<$ 7.79 & $<$ 161 & $<$ 157 \nl
1352$-$2242 & $<$ 7.53 & $<$ 263 & $<$ 269 \nl
\enddata
\tablenotetext{a}{The 90\% limit (in units of eV) on the 
equivalent width of the neutral Fe K line measured in 
the source rest-frame, with $\sigma$ = 0 eV and E = 6.4 keV} 
\tablenotetext{b}{Same as above, but now for an ionized H--like iron line 
(E = 6.97 keV)}
\end{deluxetable}

\clearpage

\begin{deluxetable}{ccccccc}
\small
\tablecaption{Power law Spectral Fits with archival ROSAT PSPC data. \label{tab5}}
\tablewidth{0pt}
\tablehead
{
\colhead{Object} & \colhead{Count Rate} & \colhead{Exposure} & 
\colhead{$N_{\rm H}$} & \colhead{$\Gamma$} & \colhead{$\chi^{2}$/dof} 
& \colhead{$F_{0.1-2 keV}$\tablenotemark{a}}\\
\colhead{} & \colhead{} & \colhead{(ks)} & \colhead{(10$^{20}$ cm$^{-2}$)} & \colhead{} & \colhead{} & \colhead{}\nl
}
\startdata
0040$+$0034$^{\dagger}$ & 0.135 & 10.7 & $\equiv$$N_{{\rm H}_{\rm gal}}$ & 2.80$\pm{0.16}$ & 8.75/6 & 11 \nl
 & & & $<$ 8.15 & 2.80$^{+0.35}_{-0.16}$ & 8.75/5 & \nl
\tableline
1101$-$264 & 0.04 & 5.12 & $\equiv$$N_{{\rm H}_{\rm gal}}$ & 1.95$^{+0.35}_{-0.40}$ & 9.9/8 & 4.1 \nl
 & & & $<$ 98.7 & 1.95 frozen & 8.9/8 & \nl
\tableline
1255$+$3536 & 0.11 & 4.06 & $\equiv$$N_{{\rm H}_{\rm gal}}$ & 2.09$^{+0.13}_{-0.14}$ & 10.8/10 & 8.6 \nl
 & & & $<$ 18.7 & 2.23$^{+0.54}_{-0.27}$ & 10.6/9 & \nl
\enddata
\tablecomments{Intervals are at 90\% confidence for one interesting parameter ($\Delta\chi^{2}$ = 2.71)\\
$^\dagger$ The quasar is offset by $\sim$ 44$\arcmin$ from the center of ROSAT PSPC observation}.
\tablenotetext{a}{Observed flux in units of 10$^{-13}$ erg cm$^{-2}$ s$^{-1}$}
\end{deluxetable}

\clearpage

\newpage
{
\centerline{\hbox{
\psfig{figure=fig1a.eps,height=7.5cm,width=7.5cm,angle=-90}
\psfig{figure=fig1b.eps,height=7.5cm,width=7.5cm,angle=-90}
}}
\vskip 1.2cm
\centerline{\hbox{
\psfig{figure=fig1c.eps,height=7.5cm,width=7.5cm,angle=-90}
\psfig{figure=fig1d.eps,height=7.5cm,width=7.5cm,angle=-90}
}}
\centerline{\hbox{
\psfig{figure=fig1e.eps,height=7.5cm,width=7.5cm,angle=-90}
\psfig{figure=fig1f.eps,height=7.5cm,width=7.5cm,angle=-90}
}}
\vskip 1.2cm
\centerline{\hbox{
\psfig{figure=fig1g.eps,height=7.5cm,width=7.5cm,angle=-90}
\psfig{figure=fig1h.eps,height=7.5cm,width=7.5cm,angle=-90}
}}
\centerline{\hbox{
\psfig{figure=fig1i.eps,height=7.5cm,width=7.5cm,angle=-90}
\psfig{figure=fig1l.eps,height=7.5cm,width=7.5cm,angle=-90}
}}}

\clearpage
\begin{figure*}
\plottwo{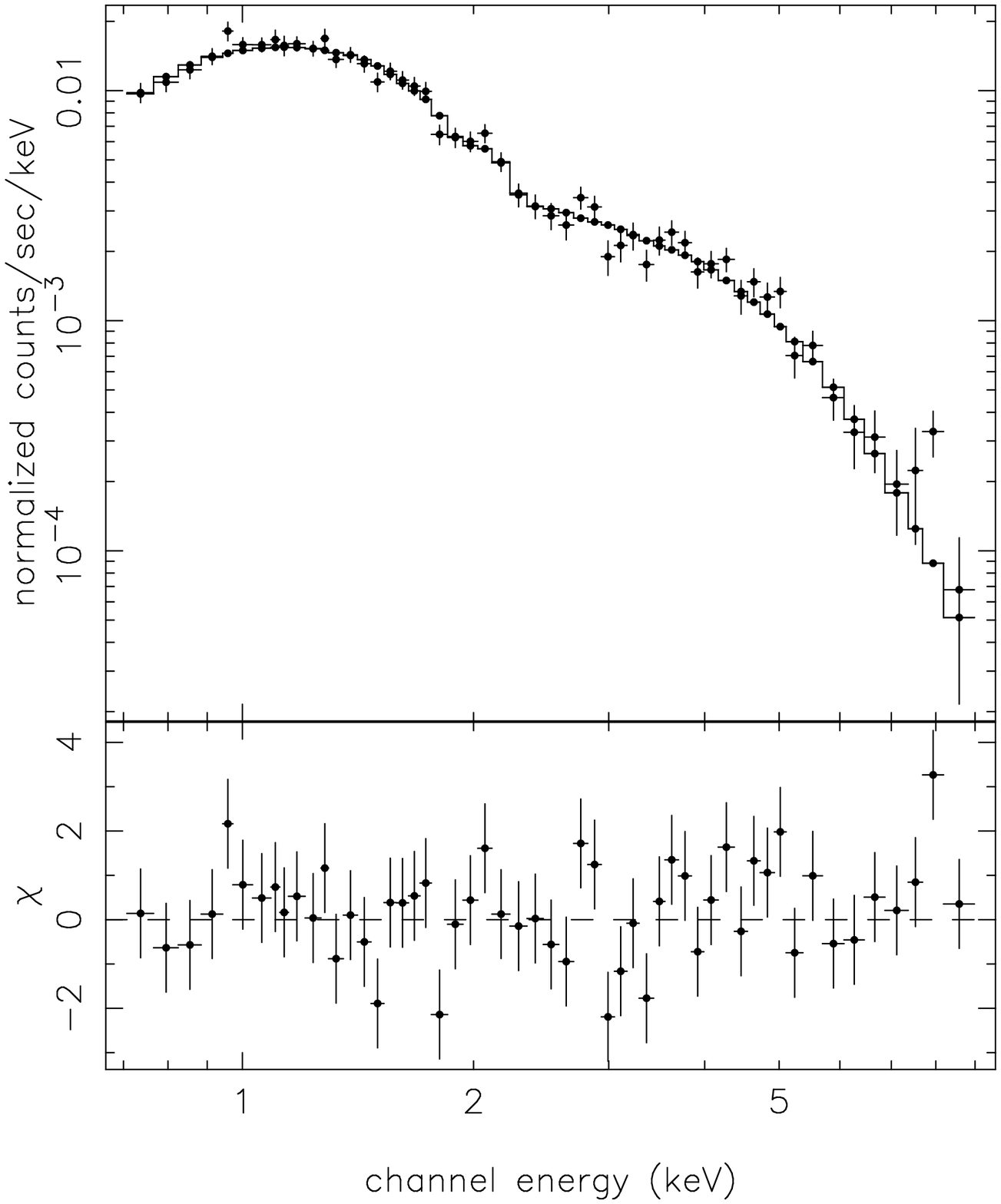}{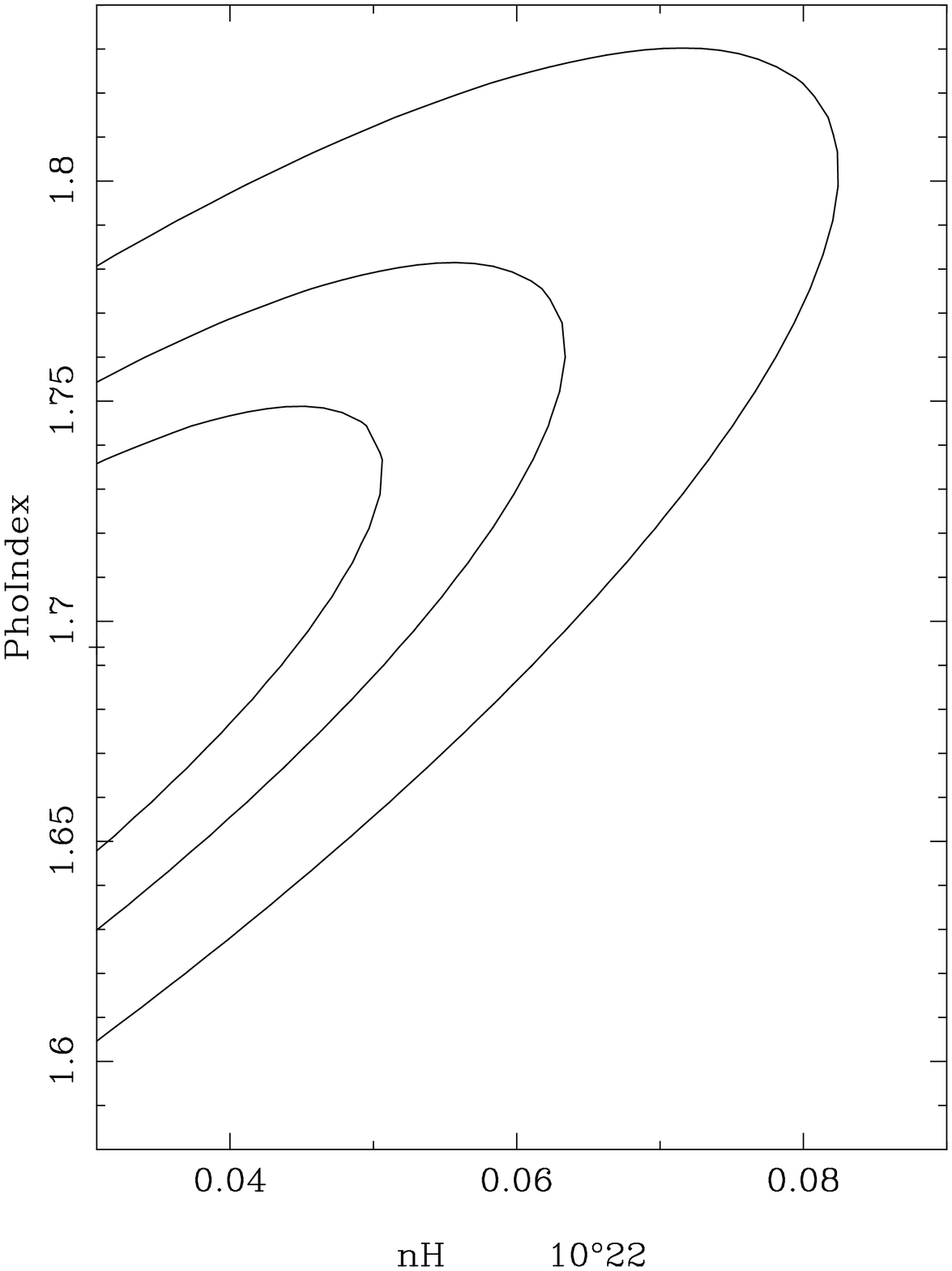}
\end{figure*}

\clearpage
\begin{figure*}
\plottwo{fig3a.eps}{fig3b.eps}
\end{figure*}

\clearpage
\begin{figure}
\plotone{fig4.eps}
\end{figure}

\clearpage
\begin{figure}
\plotone{fig5.eps}
\end{figure}


\begin{references}

\reference{}
Anders, E., \& Grevesse, N. 1989, Geochimica et Cosmochimica Acta, 53, 197
\reference{}
Arnaud, K. A. 1996, in ASP Conf. Ser., Vol.~101, Astronomical Data Analysis Software 
and Systems V, ed. G. Jacoby \& J. Barnes (San Francisco: ASP), 17 
\reference{}
Avni, Y. 1976, \apj, 210, 642
\reference{}
Balucinska-Church, M., \& McCammon, D. 1992, \apj, 400, 699
\reference{}
Bechtold, J., et al. 1994, \aj, 108, 759
\reference{}
Brinkmann, W. 1992, in MPE Report 235, X--ray Emission from AGN 
and the Cosmic X--ray Background, ed. W. Brinkmann, 
\& J. Tr\"{u}mper (MPE: Garching), 143
\reference{}
Brunner, H., Friedrich, P., Zimmermann, H.-U., \& Staubert, R. 1992, 
in MPE Report 235, X--ray Emission from AGN 
and the Cosmic X--ray Background, ed. W. Brinkmann, 
\& J. Tr\"{u}mper (MPE: Garching), 198
\reference{}
Canizares, C. R., \& White, J. L. 1989, \apj, 339, 27
\reference{}
Cappi, M., Matsuoka, M., Comastri, A., Brinkmann, W., Elvis, M., Palumbo, 
G. G. C., \& Vignali, C. 1997, \apj, 478, 492
\reference{}
Cappi, M., Matsuoka, M., Otani, C., \& Leighly, K. 1998, \pasj, 50, 213
\reference{}
Comastri, A., Setti, G., Zamorani, G., Elvis, M., Giommi, P., Wilkes, B. J., \& 
McDowell, J. C. 1992, \apj, 384, 62
\reference{}
Comastri, A., Setti, G., Zamorani, G., \& Hasinger, G. 1995, \aap, 296, 1
\reference{}
Dickey, J. M., \& Lockman, F. J. 1990, \araa, 28, 215
\reference{}
Elvis, M., Fiore, F., Wilkes, B. J., McDowell, J. C., \& Bechtold, J. 1994, 
\apj, 422, 60
\reference{}
Elvis, M. 1996, in MPE Report 263, R\"{o}ntgenstrahlung from the Universe, ed. H.U. 
Zimmermann, J. Tr\"{u}mper, \& H. Yorke (MPE: Garching), 409
\reference{}
Fiore, F., Elvis, M., Giommi, P., \& Padovani, P. 1998, \apj, 492, 79
\reference{}
Gendreau, K. C. 1995, PhD Thesis, Massachussets Institute of Technology
\reference{}
Gendreau, K. C., et al. 1995, \pasj, 47, L5
\reference{}
Hasinger, G., Burg, R., Giacconi, R., Hartner, G., Schmidt, M., Tr\"{u}mper, J., 
\& Zamorani, G. 1993, \aap, 275, 1
\reference{}
Iwasawa, K., \& Taniguchi, Y. 1993, \apj, 413, L15
\reference{}
Kii, T., et al. 1991, \apj, 367, 455 
\reference{}
Kriss, G. A., \& Canizares, C. R. 1985, \apj, 297, 177
\reference{}
Laor, A., Fiore, F., Elvis, M., Wilkes, B. J., \& McDowell, J. C. 1997, 
\apj, 477, 93
\reference{}
Lawson, A. J., Turner, M. J. L., Williams, O. R., Stewart, G. C., \& Saxton, R. D. 
1992, \mnras, 259, 743
\reference{}
Lawson, A. J., \& Turner, M. J. L. 1997, \mnras, 288, 920
\reference{}
Makishima, K., et al. 1996, \pasj, 48, 171
\reference{}
Marshall, F. E., Boldt, E. A., Holt, S. S., Miller, R. B., Mushotzky, R. F., 
Rose, L. A., Rothschild, R. E., \& Serlemitsos, P. J. 1980, \apj, 235, 4
\reference{}
Mushotzky, R. F., Done, C., \& Pounds, K. A. 1993, \araa, 31, 717
\reference{}
Nandra, K., \& Pounds, K. A. 1994, \mnras, 268, 405
\reference{}
Nandra, K., George, I. M., Turner, T. J., \& Fukazawa, Y. 1996, \apj, 464, 165
\reference{}
Nandra, K., George, I. M., Mushotzky, R. F., Turner, T. J., \& Yaqoob, T. 1997a, 
\apj, 477, 602
\reference{}
Nandra, K., George, I. M., Mushotzky, R. F., Turner, T. J., \& Yaqoob, T. 1997b,
\apjl, 488, L91
\reference{}
Osterbrock, D. E. 1991, Rep. Prog. Phys., 54, 579
\reference{}
Otani, C., \& Dotani, T. 1994, ASCA Newsl., 2, 25
\reference{}
Reeves, J. N., Turner, M. J. L., Ohashi, T., \& Kii, T. 1997, \mnras, 292, 468
\reference{}
Siebert, J., Matsuoka, M., Brinkmann, W., Cappi, M., Mihara, T., 
\& Takahashi, T. 1996, \aap, 307, 8
\reference{}
Tanaka, Y., Inoue, H., \& Holt, S. S. 1994, \pasj, 46, L37
\reference{}
Turner, T. J., \& Pounds, K. A. 1989, \mnras, 240, 833
\reference{}
V\'eron-Cetty, M. P., \& V\'eron, P. 1996, ESO Sci. Rep., 17, 1
\reference{}
Voges, W., et al. 1996, IAU Circ., 6420, 2
\reference{}
Voges, W., Englhauser, J., Gruber, R., Haberl, F., K$\ddot{u}$rster, 
M., Pietsch, W., \& Zimmermann, H. U. 1994, MPE ROSAT News, 32
\reference{}
White, N. E., Giommi, P., \& Angelini, L. 1994, \baas, 185, 41.11
\reference{}
Wilkes, B. J., \& Elvis, M. 1987, \apj, 323, 243
\reference{}
Williams, O. R., et al. 1992, \apj, 389, 157
\reference{}
Worrall, D. M., Tanambaum, H., Giommi, P., \& Zamorani, G. 
1987, \apj, 313, 596
\reference{}
Yamashita, A., Matsumoto, C., Ishida, M., Inoue, H., Kii, T., Makishima, K., 
Takahashi, T., \& Tashiro, M. 1997, \apj, 486, 763
\reference{}
Yaqoob, T. 1998, \apj, 500, 893
\reference{}
Yuan, W., Brinkmann, W., Siebert, J., \& Voges, W. 1998, \aap, 330, 108
\reference{}
Zamorani, G., et al. 1981, \apj, 245, 357
\reference{}
\.{Z}ycki, P. T., \& Czerny, B. 1994, \mnras, 266, 653


\end{references}
\end{document}